\documentstyle[twocolumn,aps]{revtex}
\def\la{\mathrel{\vcenter{\offinterlineskip\halign{\hfil
$\displaystyle##$\hfill\cr<\cr\sim\cr}}}}
\def\ga{\mathrel{\vcenter{\offinterlineskip\halign{\hfil
$\displaystyle##$\hfill\cr>\cr\sim\cr}}}}

\def\omegabar{\hbox{${\bar{\omega}}$}}
\def\boldom{\hbox{\pmb{$\omega$}}}

\def\pmb#1{\setbox0=\hbox{#1}
   \kern-0.025em\copy0\kern-\wd0
   \kern.05em\copy0\kern-\wd0
   \kern-.025em\raise.0433em\box0}

\begin{document}

\draft
\preprint{UF-IFT-HEP-97-2/6;~astro-ph/9702014\\}

\title{Relativistic electrons on a
rotating spherical magnetic dipole:~surface orbitals\\
}

\author{James M. Gelb\cite{gelb}\\}
\address{
Department of Physics\\
University of Texas at Arlington\\
Arlington, Texas 76019\\
}
\author{Kaundinya S. Gopinath and Dallas C. Kennedy\cite{kennedy}\\}
\address{
Department of Physics\\
University of Florida at Gainesville\\
Gainesville, Florida 32611\\
}

\date{\today}
\maketitle

\begin{abstract}
The semiclassical orbitals of a relativistic
electron on a rotating sphere threaded by an intense
magnetic dipole field are examined.  Several physically distinct regimes 
emerge, depending on the relative sizes of the mass, total energy, canonical 
azimuthal angular momentum,
and magnetic field strength.  Magnetic flux enclosed by orbits is quantized
very close to the poles, suggesting a quantum Hall-like state. Application 
of this system to neutron star surfaces is outlined.
\end{abstract}


\narrowtext

\section*{1. INTRODUCTION}
Neutron star pulsars are known to have surface magnetic fields of up to 
10$^{12-13}$ Gauss~\cite{long92,mich91,shap83}.
The magnetic field lines dragged in by neutron star collapse are presumably 
squeezed by magnetic 
flux conservation in the contracting plasma.  However, that assumption does
not answer the question of how the magnetic field sustains itself at later
times.  The crust, and at least part of the interior of a neutron star,
have electrons and protons.  Their internal macroscopic currents
might also affect the field, if not actually create it, but these internal 
currents are in turn
strongly affected by the intense field.  We assume  that such currents are
essentially electronic.  If the density of electrons is
low and/or the magnetic field large, a phenomenon akin to the quantum Hall
effect\cite{frad91} should be expected.

We select a simplified system retaining some features of a 
realistic neutron star.  This system is chosen to isolate quantum Hall-like
surface states: a charged particle constrained to move on the surface of a 
sphere of radius $R$ threaded by an intense magnetic dipole.  The
treatment  is restricted to {\it surface} states as their distinctive
properties may someday be observable.  The sphere rotates with angular 
velocity $\mathbf{\Omega},$ not necessarily parallel to the magnetic 
axis.  Our main discussion assumes that the two axes are parallel in
order to estimate the effect of rotation.  This is followed by the general
case of tilted axes, which is not greatly different for a moderate rotation rate.
The problem can be treated relativistically by use of the appropriate metric 
for the rotating frame.  The surface orbits serve as a prelude to
the three-dimensional tilted, rotating system, which we treat in a separate 
paper~\cite{gelb98}.

Neglecting rotation, the distinct physical regimes can be characterized by 
comparison of three dimensionless parameters:
\begin{eqnarray}
\beta_0 & \equiv & |q|B_0R/(2mc^2)\quad , \nonumber \\
\epsilon & \equiv & E/mc^2\equiv \beta_0\eta\quad , \nonumber \\
l_\phi & \equiv & 2cP_\phi /(qB_0R^2) \nonumber
\end{eqnarray}
which are, respectively, the magnetic field strength $B_0$ in rescaled units at the 
magnetic poles; particle energy $E$ in units of the rest mass; and 
particle {\it canonical} 
azimuthal angular momentum $P_\phi$ in rescaled units about the rotation axis.
For electrons on a neutron star with the strongest measured fields,
$\beta_0\simeq$ (0.1)$R/(2\lambda_C)\sim$ 2.5$\times10^{15},$ where 
$\lambda_C$ is the electron Compton wavelength = $2\pi\hbar /(m_ec).$
The relevant energy scale is then set not by $mc^2,$ 
but by $\eta$ = $\epsilon/\beta_0.$

The regime of $\eta\ga$ 1
is the {\it ultrahigh energy} case, $E\simeq$ 10$^{21}$ eV or higher,
depending on $B_0$ and $R$ but not on $m.$  
The magnetic field $\beta_0$ sets the scale for $P_\phi :$
when $|l_\phi|\sim$ 1, the {\it ultrahigh $P_\phi$} case, the 
effect of $P_\phi$ can overcome the inhibiting effect of the field.
When $\eta$ and $l_\phi$ are small, the charged particle has no 
allowed region of motion on the spherical surface except very close to the 
rotational and magnetic poles.  A large $l_\phi$ allows narrow regions away 
from the poles, but only $\eta\ga$ 1 allows charged particle motion
over a substantial portion of the sphere.

The rotational angular velocity is rescaled to $\omegabar$ = $\Omega R/c,$ 
which is $\la$ 0.1 in realistic cases~\cite{long92,mich91,shap83}.
We take $\omegabar$ = 0.1 throughout as illustrative, being
an order-of-magnitude upper limit on pulsar rotation~\cite{lyne90}.

The simplest treatments, in the limit of infinite field strength, of charged 
particles trapped on spheres by intense magnetic fields result in particles 
frozen in place in the crust~\cite{shap83,rude75}.
The treatment here adds the feature of expanding the 
particle motion in inverse powers of the field strength.  Although electrons
are stripped from the neutron star surface by the rotation-induced electric
field, the bulk of electrons remain in the crust to prevent significant charge
separation,
with the surface sheathed by a thin space charge.  The surface 
space charge is stabilized by the Coulomb force (with the positive crystal)
opposing the induced electric field.  Only a small fraction of electrons
are accelerated into the stellar wind~\cite{mich91}. The
application of this paper is to quantum single- and many-body states
where radiation emission is neglected.  This is exact for charged particles
in their ground states or in excited one-body states unable to decay by Pauli 
exclusion blocking in the presence of other fermions.  We also seek a general 
classification of possible orbitals, based 
on the kinematic parameters $\epsilon$, $\eta$, and $l_\phi$ and rotation
$\omegabar$.\\
 
\section*{2. CLASSICAL KINEMATICS}

We explicitly show all factors of $c$ and, in Sect.~IV, of $\hbar .$
Thus $x_\mu$ = $(x_0, {\bf x})$ = $(ct, {\bf x}).$  The metric has dimensions
with signature $(+---).$

\subsection*{2.1 Generally covariant Lagrangian}

The general relativistic action of a particle of mass $m$ and charge 
$q$ in an external electromagnetic and gravitational field is
\begin{equation}
S=-mc\int \sqrt{g_{\mu\nu}(x)dx^{\mu}dx^{\nu}}+\frac{q}{c}\int A_{\mu}(x)dx^{
\mu}
\end{equation}
\noindent
with fixed endpoints in parameter-independent form~\cite{land75}.
The path parameter is the proper time $\tau$.  Then if
\begin{equation}
\dot{x^{\mu}}\equiv\frac{dx^{\mu}}{d\tau}\quad ,
\end{equation}
\noindent
the general relativistic Lagrangian of a particle in the external fields
$A_{\mu}$ and $g_{\mu\nu}$ is
\begin{equation}
L=-mc\sqrt{g_{\mu\nu}\dot{x^{\mu}}\dot{x^{\nu}}} + \frac{q}{c}A_{\mu}\dot{x^{
\mu}}\quad .
\end{equation}
\noindent
The canonical momenta are given by
\begin{equation}
P_{\mu}=\frac{\partial L}{\partial\dot{x^{\mu}}}=\frac{-mc g_{\mu\nu} \dot{x^{\nu}}}{\sqrt{g_{\alpha\beta}\dot{x^{\alpha}}\dot{x^{\beta}}}} +\frac{q}{c}A_{\mu}\quad .
\end{equation}
\noindent
Using the identity
\begin{equation}
g_{\mu\nu}g^{\mu\lambda}=\delta_{\nu}^{\lambda}\quad ,
\end{equation}
\noindent
Eq.~(4) gives the constraint
\begin{equation}
g^{\mu\nu}(P_{\mu}-\frac{q}{c}A_{\mu})(P_{\nu}-\frac{q}{c}A_{\nu})=(mc)^{2}\quad .
\end{equation}
\noindent
The equations of motion are given by
\begin{equation}
\frac{\partial L}{\partial x^{\mu}}-\frac{dP_{\mu}}{d\tau}=0\quad .
\end{equation}

\subsection*{2.2 Two-dimensional rotating sphere}

We choose axes so that the magnetic dipole is along the $\theta = 0$ 
direction.  The threading dipole magnetic field has polar strength $B_{0}:$ 
\begin{eqnarray}
A_{\theta} & = & 0 \\
A_{r} & = & 0 \\
A_{\phi} & = & (B_{0}R^2/2)\sin^2 \theta~,
\end{eqnarray}
\noindent 
where $A_\phi$ is defined in the rotating spherical coordinates.

The rotation axis is tilted at an angle $\theta_0$ with respect to the magnetic
dipole in the $\phi$ = 0, $\pi$ plane (Fig.~\ref{fig1}). 
(This apparently strange
choice of coordinates is motivated by the the fact that the magnetic
field still dominates the rotational effects.)  The metric in a spherical 
polar coordinate system $(r,\theta ,\phi )$ rotating with the sphere is 
given by the line element
\begin{eqnarray}
ds^{2} &=& g_{\mu\nu}dx^{\mu}dx^{\nu}\nonumber\\
 &=&c^{2}(1-\boldom^2)dt^{2}
-r^{2}(d\theta^{2}+\sin^{2}\theta d\phi^{2}) - \nonumber \\
 & & 2cr\omega_\phi\sin\theta~dt~d\phi - 2cr\omega_\theta~dt~d\theta\quad .
\end{eqnarray}
The vector $\boldom$ is defined from the rotational angular velocity vector
$\mathbf{\Omega}$ by $\boldom$ = $\mbox{\boldmath $\Omega\times r$}/c$,
with components
\begin{eqnarray}
\omega_{\phi} & = & \omegabar [\cos\theta_0\sin\theta - 
\sin\theta_0\cos\theta\cos\phi ]\quad , \nonumber \\
\omega_{\theta} & = & -\omegabar\sin\theta_0\sin\phi\quad , \\
{\boldom}^2 & = & \omega^2_\phi + \omega^2_\theta\quad . \nonumber
\end{eqnarray}
These are the appropriate generalizations
to the 
case $\theta_0\neq$ 0~\cite{land75}.

\begin{figure}
\vspace{11pc} 
\caption{
The sphere of radius $R$, threaded by a magnetic dipole field {\bf M}, and  rotating with angular velocity \mbox{\boldmath 
${\Omega}$} which is tilted by angle $\theta_0$ with respect to dipole.}
\label{fig1}
\end{figure}

The Lagrangian is expressed in terms of the proper time $\tau ,$ 
with $\dot{x^\mu}\equiv$ $dx^\mu /d\tau .$  Eq.~(7) are three 
equations, one for each momentum component, suppressing radial motion.
Since the Lagrangian does not explicitly depend on $t,$ we have
\begin{equation}
\frac{\partial L}{\partial t}=0\quad ,   
\end{equation}
\noindent
which implies
\begin{equation}
\frac{dP_{0}}{d\tau}=0\quad .
\end{equation}
\noindent
Because
\begin{equation}
d\tau=\frac{dt\sqrt{g_{\mu\nu}(dx^{\mu}/dt)(dx^{\nu}/dt)}}{c}\quad ,
\end{equation}
\noindent
we have also
\begin{equation}
\frac{dP_{0}}{dt}=0\quad .
\end{equation}
\noindent
Thus the energy $E$ = $P_0$ is conserved.
The equations of motion for $\phi$ and $\theta$ are non-trivial:
\begin{equation}
\frac{\partial L}{\partial \phi}-\frac{dP_{\phi}}{d\tau}=0\quad
\end{equation}
\noindent and
\begin{equation}
\frac{\partial L}{\partial \theta}-\frac{dP_{\theta}}{d\tau}=0\quad .
\end{equation}
\noindent

The constraint~(6) is
\begin{eqnarray}
g^{00}P_{0}^{2} + 2g^{0i}P_{0}(P_{i}-
\frac{q}{c}A_{i}) \nonumber \\ +g^{ij}(P_{i}-
\frac{q}{c}A_{i})(P_{j}-\frac{q}{c}A_{j}) \nonumber \\ =(mc)^2~,
\end{eqnarray}
\noindent
including both $g^{0\theta}$ and $g^{0\phi}$ terms.
The contravariant metric components are
\begin{eqnarray}
g^{00} = 1/c^2~, & g^{rr} = -1/r^2~, \nonumber \\
g^{\theta\theta} = -(1-\omega_{\theta}^2)/r^2~, &
\ \ g^{\phi\phi} = -(1-\omega_{\phi}^2)/(r^2\sin^2\theta)~, \\
g^{0\phi} = -\omega_\phi/(cr\sin\theta)~, &
g^{0\theta} = -\omega_\theta/(cr)~, \nonumber
\end{eqnarray}
\noindent
where $r$ = $R$ for our case.  Along the actual worldpath in spacetime,
the constraint $g_{\mu\nu}\dot{x^\mu}\dot{x^\nu}$ = $c^2$ obtains;
this condition is valid after varying the action and simplifies the equations 
of motion.
 
\section*{3. CLASSICAL ANALYSIS: ZERO TILT}

A simplified treatment of rotation assumes zero tilt, $\theta_0$ = 0.
Eq.~(8-10) define the dipole field, with $A_\phi$ defined in the 
rotating 
spherical coordinates.  If $\theta_0$ = 0, the Lagrangian does not explicitly
depend on $\phi,$ and 
\begin{eqnarray}
{dP_\phi\over dt} = 0\quad ,
\end{eqnarray}
yielding another constant of motion, the canonical azimuthal angular
momentum $P_\phi$, along with $E$.  Further,
we define
\begin{eqnarray}
l_{\theta} & = & 2cP_\theta /(qB_0R^2) \nonumber \\
 & = & \bigl( 1/\beta_0\bigr)\bigl\lbrace -1+\epsilon^{2}-2\omegabar
\epsilon\beta_0[l_{\phi}-\sin^2\theta ] -\nonumber   \\
 & & [1/\sin{^2}\theta-\omegabar^{2}]\beta_0^2[l_{\phi}-
\sin^2\theta ]^{2}\bigr\rbrace^{1/2}\quad .
\end{eqnarray}
\noindent
The opposite sign of the square root, not shown, is also valid.
We consider Eq.~(22) in three different limits of physical 
interest, with $q >$ 0.  The $q <$ 0 case can be obtained
from the $q > 0$ case by reversing the sign of $l_\phi$ and $\omegabar .$

While the $\phi (t)$ motion is trivial, the polar motion $\theta (t)$ is not.
The two angular motions, both periodic, decouple from one another because 
$l_\phi$ is conserved.  Without radial motion, the polar motion alone is a 
closed one-dimensional system.  
The two periods, $\tau_\phi$ and $\tau_\theta ,$
are in general not equal or even commensurate: their ratio
$\tau_\theta /\tau_\phi$ is not necessarily a rational number $k_\phi /
k_\theta ,$ where $(k_\phi ,k_\theta)$ are a pair of integers with no common 
divisors.
If the periods are commensurate, then the orbits can be arbitrarily
complex, but close after a time $\tau_{\rm closure}$ = $k_\theta \tau_\theta$ =
$k_\phi \tau_\phi ;$  otherwise the orbits never close.  We compute below
the magnetic flux $\Phi$ enclosed by a pole orbit (see
III.C), but the flux is well-defined only if the orbit is closed.  For 
very large fields, nonetheless, the variation of $\theta$ is ${\cal O}
(\epsilon /\beta_0)$ and tiny in these two cases (ultrahigh $P_\phi$ and 
localized pole orbits), and we define the 
enclosed 
flux by one complete revolution of $\phi$ from 0 to $2\pi$ at the approximately
constant polar angle $\theta_{\rm mid}:$
\begin{eqnarray}
\Phi (\theta_{\rm mid}) & = & \pi B_0R^2~[1 - \cos 2\theta_{\rm mid}] \nonumber
\\
& = & \Bigl(\frac{2\pi\hbar c}{|q|}\Bigr)\Big(\frac{mcR}{\hbar}\Bigr)\beta_0~[
1 - \cos 2\theta_{\rm mid}]\quad .
\end{eqnarray}
In general, this flux is macroscopically large, a product of two large 
dimensionless factors and the small elementary flux quantum.

In a uniform field, the particle's cyclotron radius and magnetic flux enclosed
are constant.  The enclosed magnetic flux is still conserved, as an adiabatic 
invariant, for slowly-varying fields~\cite{long92,land75}.
A charged particle on a sphere in the present configuration would,
in general, see a rapidly varying field.  But if the field is intense
$(\beta_0\rightarrow\infty )$ and the energy small $(\eta\ll$ 1), the 
variation of the particle's orbit from constant $\theta$  is higher order in
$1/\beta_0,$ and the enclosed magnetic flux is quasi-invariant (see also
Sect.~V).

\subsection*{3.1 Ultrahigh energy orbits}

In the limit of very high energy, i.e., $\epsilon\gg 1,$ 
Eq.~(22) becomes
\begin{eqnarray}
l_\theta = \sqrt{\eta^2 +
2\omegabar\eta\sin^2\theta -
[1 -\omegabar^2\sin^2\theta]\sin^2\theta}\quad ,
\end{eqnarray}
\noindent where $P_\phi$ has been neglected.  That is, $\eta\sim$ ${\cal O}(1)$
and $|l_\phi|\ll$ 1.  In Fig.~\ref{fig2}, we plot the right 
hand side versus $\cos\theta$ and various 
values of $E.$  We find that, for values of energy $\eta <$ a critical value 
$\eta_c(\omegabar ),$ there are 
four turning points enclosing two distinct allowed regions for the particle 
between the poles and the equator, with two of the four turning points very 
near the poles.  (These
polar turning points are nonzero because of the centrifugal effect of
$P_\phi$ and vanish as $P_\phi$ vanishes.)  As we increase the energy, the 
allowed regions expand and, for $\eta >$ $\eta_c,$ merge into a single region 
occupying essentially the whole surface of the sphere.  In the latter case,
two of the turning points merge and disappear.  The turning points in the 
ultrahigh energy limit are:
\begin{eqnarray}
\sin^2\theta_\pm = \frac{1-2\omegabar\eta \pm \sqrt{1-4\omegabar\eta}}
{2\omegabar^2}\quad .
\end{eqnarray}
Only one of these roots, $\sin^2\theta_-,$ is physical in the limit $\Omega$
$\rightarrow$ 0: $\sin^2\theta_-\rightarrow$ $\eta^2.$  If $\Omega$ = 0,
then $\eta_c$ = 1.

\begin{figure} 
\vspace{11pc}
\caption{Ultrahigh energy orbits: $P_\theta (\theta)$ for energies $\eta$ 
= 0.5 $-$ 4.0, in units of $(\beta_0\eta )(mcR).$}
\label{fig2}
\end{figure}

\subsection*{3.2 Ultrahigh $P_\phi$ orbits}

Taking $|l_\phi |$ in this case to be $\sim{\cal O}(1)$, Eq.~(22) can be 
re-expressed as
\begin{eqnarray}
l_\theta = \bigl( 1/\beta_0\bigr)\bigl\lbrace -1+\epsilon^2 
 -2\omegabar\eta\beta_0[l_{\phi} - \sin^2\theta ] - \nonumber \\
\quad [1/\sin^{2}\theta-\omegabar^{2}]\beta_0^{2}[l_{\phi}-
\sin^2\theta ]^{2}\bigr\rbrace^{1/2}\quad .
\end{eqnarray}
\noindent
The ultrahigh magnetic field introduces a 
very large negative quantity into the square root.  We expand results in
inverse powers of $\beta_0.$  For a given $P_{\phi}$ or $l_\phi ,$ the two 
allowed regions are very narrow in $\theta$ and, to ${\cal O}(1)$ in 
$1/\beta_0,$ given by the two distinct values of $\overline\theta :$
\begin{eqnarray}
\sin\overline{\theta} = \sqrt{l_\phi}\quad ,
\end{eqnarray}
\noindent
where $\overline\theta$ is defined over the range 0 to $\pi ,$ and 
$\eta\ll$ 1; that is, $\epsilon$ has been neglected compared to $l_\phi .$
Note that 0 $\le l_\phi\le$ 1.
To ${\cal O}(1/\beta_0),$ the endpoints of the allowed regions are given by
\begin{eqnarray}
\cos2\theta=\cos2\overline\theta + \delta_{1,2}/\beta_0\quad ,
\end{eqnarray}
where
\begin{eqnarray}
\delta_{1,2} = \frac{ -\omegabar\epsilon \mp [(\omegabar
\epsilon)^{2}+[1/l_\phi - \omegabar^{2}]
(\epsilon^{2}-1)]^{1/2}}{2[1/l_\phi - \omegabar^{2}]}\quad ,
\end{eqnarray}
\noindent
where the subscripts $(1,2)$ here refer to the plus-minus of the square root.
The angular width of the allowed region is thus ${\cal O}(1/\beta_0).$
There are four endpoints in total, two for each allowed region.  
Fig.~\ref{fig3} shows $\overline\theta$ as a function of $l_\phi.$  
The values of $l_{\phi}$ are restricted by the condition 0 $\le |l_\phi |\le$
$1.$

\begin{figure}
\vspace{11pc}
\caption{Ultrahigh $P_\phi$ orbits: two allowed annuli with central angles
of $\overline\theta$ (solid) and $\pi - \overline\theta$ (dashed) as functions
of $l_\phi .$}
\label{fig3}
\end{figure}

\subsection*{3.3 Localized pole orbits}

Eq.~(22) can also be written, for $P_\phi\neq$ 0, as 
\begin{eqnarray}
l_\theta & = & \big( 1/\beta_0\bigr)\bigl\lbrace -1+\epsilon^{2}-
2\omegabar\epsilon\beta_0[l_\phi-\sin^2\theta ] -\nonumber \\
& & [1/\sin^{2}\theta-\omegabar^{2}]\beta_0[l_\phi-
\sin^2\theta ]^{2}\bigr\rbrace^{1/2}\quad .
\end{eqnarray}
\noindent
For $\eta$ and $|l_\phi |\ll$ 1, allowed regions appear only if $\theta$
$\rightarrow$ 0 or $\pi .$  This case can be understood by extrapolating
the ultrahigh energy and $P_\phi$ case to $\eta$ and $l_\phi\rightarrow$ 0.  
The allowed regions are two 
narrow azimuthal annuli centered about the north and south poles 
(Fig.~\ref{fig4}).

\begin{figure}
\vspace{11pc}
\caption{Localized pole orbits: narrow allowed annuli (darkened circles) near 
poles $\theta$ = 0 and $\pi ,$ shown on sphere of radius $R.$  Magnetic dipole 
${\bf M}$ and rotational ${\bf\Omega}$ axes are parallel in this case.} 
\label{fig4}
\end{figure}

The turning points near the poles are given by
\begin{eqnarray}
\lefteqn{\theta =\bar{\theta}~\pm~\Delta\theta /2} \nonumber \\
 & & =\sqrt{l_{\phi}}~\Bigl[~1\pm \frac{1}{2}\sqrt{
[\epsilon^2-1]/(\beta^2_0l_\phi)}~\Bigr]\quad ,\quad (l_\phi\ge 0)
\end{eqnarray}
\noindent
and the same replacing $\theta$ by $\pi -\theta .$  In both cases, as the field
grows, the turning points approach zero angle with the rotation axis
and the annular width vanishes.  (The ${\cal O}(l_\phi )$ effects are too
small to show in Fig.~\ref{fig4}.)
The magnetic flux enclosed by an orbit at the rotational poles is given by
\begin{eqnarray}
\Phi (\theta\simeq 0,\pi ) & = & 2\pi R^2 B_0~{\overline\theta}^2 \nonumber \\
 & = & \Bigl(\frac{4\pi\hbar c}{|q|}\Bigr)\Bigl(\frac{mcR}{\hbar}\Bigr)\beta_0
l_\phi
\end{eqnarray}
\noindent
and the same replacing $\overline\theta$ by $\pi - \overline\theta .$
Note that $\Phi\propto$ $l_{\phi}\propto$ $P_\phi /q.$  The expression~(32)
for the magnetic flux does not have to be macroscopically large, as it consists
of two large and one small dimensionless factors times the elementary flux
quantum.  If $l_\phi$ is small enough, $\Phi$ can be microscopic, signaling
a quantum Hall-like state.\\
 
\section*{4. SEMICLASSICAL QUANTIZATION:\\ZERO TILT}
 
Although we do not carry out the full quantum analysis, a semiclassical
treatment brings out many of the desired features.  For periodic classical 
orbits, semiclassical
quantization is most easily implemented with the Wilson-Sommerfeld (W-S)
or Bohr-Sommerfeld conditions~\cite{land77}. This procedure is 
valid for $\theta$ quantization as $\theta$ motion is a complete subsystem 
alone if $\theta_0$ = 0.  The closure or non-closure of the classical orbits 
is then irrelevant.

The condition on the validity of the W-S procedure is that the particle's
de Broglie wavelength be much smaller than the length scale over which the
background field varies.  In our case,
\begin{eqnarray}
{2\pi\hbar\over{|{\bf p}|}}\ll R\quad ,\nonumber
\end{eqnarray}
\noindent a condition that holds except where $|{\bf p}|\rightarrow$ 0.  Only
infinitesimally close to the $\theta$ turning points does this
condition break down, as the three-momentum $|{\bf p}|$ is otherwise large
and $R$ enormous in any case.  The W-S semiclassical quantization is validated
by the WKB approximation, with this restriction~\cite{land77}.

The classical analysis already implies orbitals reminiscent of a
quantum lattice: quasi-free conduction bands (ultrahigh energies) and
localized states (polar and azimuthal orbitals).  The azimuthal (ultrahigh 
$P_\phi$) rings are localized in one, but not two, dimensions. They conduct 
along one direction but are trapped in the other.  The pole orbitals are
similarly localized, but are also confined in absolute position to be
near the poles.

The classically allowed regions are those in which the quantum wavefunctions
are oscillatory rather than exponential.  Because of the
the Heisenberg uncertainty principle constraint, the quantization conditions, together
with the classical spherical cyclotron relations, lead to a tightly
constrained set of semiclassical orbitals.  In a uniform, planar magnetic 
system (the Landau system), these orbitals would be equivalent to harmonic 
oscillator states~\cite{land77,bere82}.

We now impose W-S quantization.  Since $P_{\phi}$ is conserved, the azimuthal 
quantization is trivial:
\begin{eqnarray}
P_{\phi}=n_{\phi}\hbar\quad ,
\end{eqnarray}
using the Bohr form of the W-S rule, valid for circular motion.
For $P_{\phi}\gg\hbar$, $n_{\phi}$ is very large and the $\phi$ motion is
essentially classical.  We assume this to be the case, except very near
the rotational poles; in that case, the orbital size vanishes as $P_\phi
\rightarrow$ 0, so that $P_\phi$ must be small.
 
\subsection*{4.1 Localized pole states: flux quantization}

The $\theta$ motion is periodic but non-circular and requires the alternate
form of the W-S rule,
\begin{eqnarray}
2\int_{\theta_{1}}^{\theta_{2}}~P_{\theta}~d\theta=(n_{\theta}+1/2)\hbar\quad
\end{eqnarray}
for turning points $\theta_1$ and $\theta_2.$  The replacement $n\rightarrow$
$n + 1/2$ preserves the exact quantization for simple harmonic motion.
This integral is evaluated
here in a simple rectangular approximation (Simpson's rule).
With eqs.~(30,31,34), we obtain
\begin{eqnarray}
E \simeq (mc^2)\sqrt{1 + (n_{\theta}+1/2)(\hbar\beta_0/(2mcR))}\quad .
\end{eqnarray}
\noindent
Fig.~\ref{fig5} shows $E/mc^2$ as a function of $B_0$ for $n_\theta$ = 0, 1, 2.
The dimensions are set by the {\it critical field} $B_c,$ the field strength 
with $mc^2$ of energy within a Compton cube of volume $(2\pi\hbar /mc)^3:$
$B_c$ = $(m^2/\pi )\sqrt{c^5/\hbar^3}.$  Fields with this strength or greater
introduce effects of quantum field theory such as vacuum polarization.  The
energy step size is controlled by the tiny ratio of Compton wavelength $\hbar 
/mc$ to $R,$ multiplied by the enormous magnetic field $\beta_0.$  Electrons
in excited states radiate until they reach the ground state $(n_\theta$ = 0).

\begin{figure}
\vspace{11pc}
\caption{
Quantized pole states: energy $E/mc^2$ as a function of pole field 
strength $B_0$ for quantum states $n_\theta$ = 0, 1, 2.
For the electron, magnetic field units in $B_c = (m^2_e/\pi)(c^5 / 
\hbar^3)^{1/2} = 1.3\times 10^{12}$ Gauss and $q^2/(\hbar c) = 1/137$.
}
\label{fig5}
\end{figure}

The magnetic flux enclosed by a semiclassical orbit, being an adiabatic
quasi-invariant, is quantized and is given by
\begin{eqnarray}
\Phi(\theta\simeq 0,\pi ) \simeq 2\pi B_0R^2{\overline\theta}^2\quad ,
\end{eqnarray}
so that
\begin{eqnarray}
\Phi(\theta\simeq 0,\pi ) \simeq \frac{4\pi cP_\phi}{q} = \Bigl(\frac{
4\pi\hbar c}{q}\Bigr)\Bigl(\frac{mcR}{\hbar}\Bigr)\beta_0l_\phi \propto 
n_\phi\quad ,
\end{eqnarray}
analogous to the quantum Hall effect.
Any $n_\theta$ dependence in $\Phi$ is a correction to~Eq.~(37) of relative order 
${\cal O}(1/\beta_0)$ and arises from the breakdown of exact invariance of
$\Phi .$

\subsection*{4.2 Localized azimuthal rings}

In the limit of high azimuthal angular momentum $|l_\phi |\sim$ ${\cal O}(1),$
using eqs.~(26-29,34), we obtain
\begin{eqnarray}
E \simeq (mc^2)\sqrt{1 + (n_{\theta}+1/2)(\hbar\beta_0/(2mcR\sqrt{l_\phi}))}
\quad .
\end{eqnarray}
The same combination of tiny ratio multiplied by very large $\beta_0$ that 
occurred in~Eq.~(35) appears again, and excited states are again unstable
to radiation. Note, the magnetic flux enclosed by a semiclassical orbit is not
interesting in the quantum regime, as it is macroscopically large.

\subsection*{4.3 Poleward conduction: quasi-continuum}

In the limit of ultrahigh energies using eqs.~(24,25,34), we obtain 
\begin{eqnarray}
E\simeq {\cal C}(n_\theta+1/2)~(\hbar c/R)\quad ,
\end{eqnarray}
where ${\cal C}\approx$ 4, 1, or 1/2 for $\eta\la ,$ $\sim ,$ or $\ga$
1.  Note that the energies scale as harmonic oscillator energies, where the 
frequency is set by the size of the sphere, {\it not} the magnetic field.  As 
in Sect.~III.A, we have assumed $P_\phi$ to be negligibly small.  In 
realistic cases, $\hbar c/R$ $\ll mc^2$ and $E,$ so that the energy levels are
very closely spaced.
Such levels form a quasi-continuum that allows the charged particles
to conduct almost freely along the polar directions, subject only to
external crystal resistance and radiation losses.  For these ultrahigh
energies, $n_\theta$ is so large that the motion is essentially classical.
The energy levels depend only weakly on $\Omega ,$ for small $\Omega R/c$.

The magnetic flux $\Phi$ is not interesting in this case because the
allowed region is a large part or all of the sphere.  Flux quantization is
not relevant because the energy is independent of the magnetic field $B_0,$
apart from the overall constant ${\cal C}.$
Only if the energy $E$ is taken $\ll {\cal O}(|q|B_0R/2)$ do $B_0$ and $P_\phi$
become important again; this is the localized pole orbit case.

\subsection*{4.4 Density of states}

The density of states is important in any charged fermion system for
determining the conductivity. The semiclassical density of states 
$d{\cal N}/dE$ is\cite{land77}
\begin{eqnarray}
d{\cal{N}}=\frac{d\theta~dP_{\theta}~d\phi~dP_{\phi}}{(2\pi\hbar)^2}\quad ,
\end{eqnarray}
and the differential surface area element is
\begin{eqnarray}
d{\cal{S}}=R^2~\sin\theta~d\theta~d\phi\quad .
\end{eqnarray}
Thus we have for the the number of states per unit area
per unit azimuthal angular momentum per unit energy
\begin{eqnarray}
\frac{d^3{\cal{N}}}{d{\cal{S}}~dP_{\phi}~dE}=\frac{dP_{\theta}}{dE}
\Bigl(\frac{1}{4\pi^2\hbar^2R^2\sin\theta}\Bigr)\quad .
\end{eqnarray}
Eq.~(22) can be rewritten as
\begin{eqnarray}
P_{\theta}/mcR & = [(E/mc^2)^2-\quad \nonumber \\
 & (E/mc^2)a_1(\theta ;P_{\phi},\Omega )-
  a_0(\theta;P_{\phi},\Omega )]^{1/2}~,\quad
\end{eqnarray}
with
\begin{eqnarray}
a_1 & = 2(\Omega /mc^2)[P_\phi -(qB_0R^2/2c)\sin^2\theta ]~,\quad
\end{eqnarray}
\begin{eqnarray}
a_0 & = 1+ (1/\sin^2\theta - \quad\nonumber \\
 & {\left(\Omega R/c\right)^2}
[P_\phi - (qB_0R^2/2c)\sin^2\theta]^2/(mcR)^2~.\quad
\end{eqnarray}
\noindent
Therefore
\begin{eqnarray}
\frac{dP_{\theta}}{dE}=(mcR/P_{\theta})[E/mc^2-a_1/2](R/c)\quad .
\end{eqnarray}
The factor $dP_\theta /dE$ is shown as a function of $\eta$ in 
Fig.~\ref{fig6} 
for $|l_\phi |\ll$ 1, and as a function of $E/mc^2$ in 
Fig.~\ref{fig7} for 
$l_\phi\sim$ 0.1. 
For low energies, $E\rightarrow mc^2,$ this function approaches
\begin{eqnarray}
[1-a_1-a_0]^{-1/2}\bigl\lbrace 1-a_1/2-\frac{[a_0+a^2_1/4]}{1-a_1-a_0}
(K/mc^2)\bigr\rbrace~,
\end{eqnarray}
where $K$ = $E - mc^2.$  In the ultrarelativistic limit, $E\gg mc^2$ and/or
$\eta\ga$ 1, $dP_\theta /dE$ approaches a constant, $R/c,$ a characteristic
of two-dimensional Landau states.  Note the large peak in $dP_\theta /dE$
at low $\epsilon$ or $\eta ,$ the semiclassical edge of the discretized
quantum regime.

\begin{figure}
\vspace{11pc}
\caption{Density of states factor $dP_\theta /dE,$ in units of $R/c,$ as a
function of energy $\eta$ for various values of  $\theta$  and
$|l_\phi |\ll$ 1.}
\label{fig6}
\end{figure}

\begin{figure}
\vspace{11pc}
\caption{Density of states factor $dP_\theta /dE,$ in units of $R/c,$ as a 
function of energy $E/mc^2$ for $l_\phi$ = 0.1}
\label{fig7}
\end{figure}

In the quantum limit,
\begin{eqnarray}
d{\cal{N}}=dn_{\theta}(\frac{d\phi~dP_{\phi}}{2 \pi \hbar})\quad ,
\end{eqnarray}
treating $P_\phi$ as continuous.  Equation~(42) becomes:
\begin{eqnarray}
\frac{d^3{\cal{N}}}{d{\cal{L}}~dP_{\phi}~dE}=\frac{dn_\theta}{dE}
\Bigl(\frac{1}{2\pi\hbar R^2\sin\theta}\Bigr)\quad ,
\end{eqnarray}
\noindent
where $d{\cal L}$ = $R\sin\theta~d\phi ,$ a unit of azimuthal arc length.
The function $dn_\theta /dE$ can be obtained from eqs.~(34,~35,~or~38). 
For the localized pole states of Eq.~(35), take $E^2$ =
$(mc^2)^2$ + $(n_\theta + 1/2)\varepsilon^2_1,$ where $\varepsilon^2_1$ =
$\varepsilon^2_1(qB_0,R,P_\phi ).$  Then
\begin{eqnarray}
\frac{dn_\theta}{dE} = \frac{dn_\theta}{dK} = 
\frac{2E}{\varepsilon^2_1} = \frac{2(mc^2+K)}{\varepsilon^2_1}
\end{eqnarray}
for $E^2\ge$ $(mc^2)^2 + \varepsilon^2_1/2.$\\

Related to the density of states is the degeneracy factor for a given 
$n_\theta$ level.  Semiclassically, this degeneracy arises from shifting
an orbital center and is identical to the degeneracy
of planar Landau states~\cite{frad91}.  The number of degenerate states 
available per $d{\cal S}$ per $dP_\phi$ at state $n_\theta$ is
\begin{eqnarray}
\frac{d^2{\cal N}}{d{\cal S}~dP_\phi}(n_\theta ) & = \nonumber \\
 & \qquad\Bigl(\epsilon_1/4\pi\hbar^2cR\sqrt{l_\phi}\Bigr)
\sqrt{n_\theta +1/2}~.
\end{eqnarray}

\section*{5. NON-ZERO TILT: APPROXIMATE ORBITALS}

\subsection*{5.1 Adiabatic quasi-invariance}

As in Sects. II and III, the energy $E$ = $P_0$ is exactly conserved, but 
$P_\phi$
or $l_\phi$ no longer is, unless $\omegabar\sin\theta_0$ is zero.
$P_\theta$ or $l_\theta$ was never conserved, 
but in with zero tilt, its motion was
exactly separable with $l_\phi$ conserved.  That separability is also
lost if $\omegabar\sin\theta_0\neq$ 0.

When $\theta_0\neq$ 0, the non-conservation of $l_\phi$ is apparent from the
equation of motion~(10):
\begin{eqnarray}
\frac{dl_\phi}{d\tau} & = & \nonumber \\
 & \frac{{\rm sgn}(q)}{cR\beta_0}
\Big[\frac{1}{2}\frac{\partial\boldom^2}
{\partial\phi} + \frac{\partial\omega_\phi}{\partial\phi}
(cr\dot{\phi}\dot{t}\sin\theta ) + \frac{\partial\omega_\theta}{\partial\phi}
(cr\dot{\theta}\dot{t})\Big]\quad ,
\end{eqnarray}
\noindent
using the general metric for $\theta_0\neq$ 0. The non-conservation of
$l_\phi$ is ${\cal O}(1/\beta_0),$ as well as requiring $\omegabar
\sin\theta_0\neq$ 0. In the limit of ultra-intense magnetic field $\beta_0,$
the motion is not qualitatively different from the zero-tilt
case.  The orbits are more involved and possibly chaotic, but so focused
by the intense field that, to zeroth order in $(1/\beta_0),$ the results
of Sects. II and III with $\theta_0$ = 0 are still qualitatively valid.

The approximate conservation of $l_\phi$ validates a perturbative treatment
of the classical orbits.  In Sect. II, where $l_\phi$ is exactly conserved, 
the action variable
\begin{eqnarray}
J_\phi = \oint d\phi\ P_\phi
\end{eqnarray}
is an adiabatic invariant~\cite{land75,land76}.
This allows for the semiclassical quantization of $P_\phi$ as a trivial step in 
that case as well as the exactly separable treatment of $P_\theta$.
We thus expect any effects of the non-zero tilt to be suppressed by powers of
$\omegabar$ and $1/\beta_0.$

\subsection*{5.2 Classical motion}

Since the $\phi$ motion to zeroth order in $1/\beta_0$ is periodic, global
expressions involving the $\phi$ coordinate (such as $E,$ but not the
orbit $\phi (\tau ),$ $\theta (\tau )$) can be averaged over the full circle 
$\phi$ $\in [0,2\pi ]$ with $l_\phi$ treated as constant, in order to determine
the first-order corrections~\cite{land76}.
The ${\cal O}(1/\beta_0)$ 
correction to $dl_\phi/d\tau$ = 0 is obtained from averaging~(14) over $\phi :$
\begin{eqnarray}
\Big\langle\frac{dl_\phi}{d\tau}\Big\rangle_\phi = 
\frac{{\rm sgn}(q)}{mcR\beta_0}
\Big\langle\frac{\partial L}{\partial\phi}\Big\rangle_\phi = 0\quad ,
\end{eqnarray}
where all the terms are $\propto$ $\cos\phi$, $\sin\phi$, or $\cos\phi\sin\phi$
and average to zero.
Consequently, the non-conservation of $l_\phi ,$ on average, actually starts 
at ${\cal O}(1/\beta^2_0).$

We consider the same set of limiting cases as in Sect.~II.
In the limit of {\it ultrahigh energy,} $\beta_0\gg l_\phi ,$
the constraint~(19) with Eq.~(20) becomes:
\begin{eqnarray}
\lefteqn{l_\theta (1-\omega^2_{\theta})/\beta_0 = \omega_\theta~\pm} \\
 & & \sqrt{(\omega_\theta)^2 + 
(1-\omega_{\theta}^2)[\eta^2 +  2 \eta\omega_\phi\sin\theta - (1-\omega^2_
{\phi})\sin^2\theta]}~, \nonumber
\end{eqnarray}
retaining all powers of $\eta$.
Figs.~\ref{fig8} and~\ref{fig9} 
show the allowed regions for several combinations of $E$
and $\theta_0,$ where the allowed regions are defined by the requirement
that $l_\theta$ in~(54) be real.  Note that the allowed regions of 
Figs.~\ref{fig8}
and~\ref{fig9} are no longer functions of $\theta$ alone, unlike the case of
eqs.~(24,25).  The axial symmetry about the dipole is lost, but for
moderate values of $\omegabar ,$ the asymmetry is not extreme.  For large
tilts $(\theta_0\rightarrow\pi),$ the energy $\eta$ must be somewhat higher
than in the zero-tilt case to make the entire sphere allowed.  If $\phi$ 
averaging is applied to the constraint~(19), the axial symmetry about the
dipole is restored.

\begin{figure}
\vspace{22pc} 
\caption{
Allowed regions of real $l_\theta$ in the ultrahigh energy limit, for $\eta$ = 
0.5.  Surface coordinates are colatitude $\theta$ and longitude $\phi$ in
degrees.  (a) $\theta_0$ = 45$^\circ .$  (b) $\theta_0$ = 135$^\circ .$}
\label{fig8}
\end{figure}

\begin{figure}
\vspace{22pc} 
\caption{
Allowed regions of real $l_\theta$ in the ultrahigh energy limit, for $\eta$ = 
1.0.  Surface coordinates are colatitude $\theta$ and longitude $\phi$ in
degrees.  (a) $\theta_0$ = 45$^\circ .$  (b) $\theta_0$ = 135$^\circ .$}
\label{fig9}
\end{figure}

The limit where $l_\phi$ cannot be neglected, but $\eta\ll$ $l_\phi ,$ is
the {\it ultrahigh} $P_\phi$ case; $l_\theta$ is then zero except within the
narrow allowed regions.  If the tilt is zero, eqs.~(26-29), 
the particle orbits are 
almost infinitely narrow circles at constant $\theta .$  The $\theta$ motion is
${\cal O}(1/\beta_0)$ relative to the $\phi$ motion.  With non-zero tilt and
$l_\phi$ approximately conserved, the quasi-circular orbits occur at the same
angle as the no tilt case,
$\sin\overline\theta$ = $\sqrt{l_\phi},$ for $l_\phi\ge$ 
0, with ${\cal O}(1/\beta_0)$ and ${\cal O}(\omegabar\sin\theta_0/\beta_0)$ 
corrections.

The same argument holds for the {\it small angle} case, which is related to
the ultrahigh $P_\phi$ limit in the limit $l_\phi\rightarrow$ 0,
with $\eta$ still negligible.  The approximate conservation of $l_\phi$
still holds in the limit of large $\beta_0,$ with additional
corrections of relative order $\omegabar\sin\theta_0.$

The magnetic flux enclosed by an orbit is, to leading order in $1/\beta_0,$
just as in Sect.~II, eqs.~(32,36,37).  The averaged rotational 
corrections are of order ${\cal O}(\omegabar^2).$

\subsection*{5.3 Semiclassical quantization}

In the present case, the quantization of $l_\phi$ in the Wilson-Sommerfeld 
condition~(33) is only approximate, but valid through
${\cal O}(1/\beta_0).$

The {\it ultrahigh energy} case is not qualitatively different if $\theta_0
\neq$ 0.  As seen from the classical orbits, some regions of the sphere
are forbidden for large tilt and moderate $\eta .$  The quantum energy
levels are still approximately the harmonic oscillator levels
given by Eq.~(39), depending only weakly on the field.

The {\it ultrahigh} $P_\phi$ case results in the energy levels:
\begin{eqnarray}
E/mc^2 & \simeq \Big\{1 - (\omegabar^2/2)(\sin^2\theta_0 + 
l_\phi\cos^2\theta_0) +\quad \nonumber \\
 & (1-(5\omegabar^2/4)\sin^2\theta_0 +
\omegabar^2l_\phi\cos^2\theta_0 ) \times \quad \nonumber \\
 & (n_\theta + 1/2)\hbar\beta_0/(2mcR\sqrt{l_\phi})\Big\}^{1/2}~,\quad
\end{eqnarray}
to leading order in $\beta_0$ and $\omegabar ,$ with non-zero tilt.  Note that
the rotational corrections enter at ${\cal O}(\omegabar^2).$  The 
corrections multiplying $\beta_0$ can either raise or lower the energy, but 
the first correction (the pure centrifugal effect) always lowers the energy.  

The {\it localized pole} states are also corrected by rotational effects.
Their levels are:
\begin{eqnarray}
E/mc^2 \simeq \Big\{ 1 - (\omegabar^2/4)\sin^2\theta_0 +\quad \\
  (1-(3\omegabar^2/4)\sin^2\theta_0)
(n_\theta + 1/2)\hbar\beta_0/(2mcR)\Big\}^{1/2}~,\quad \nonumber
\end{eqnarray}
and reproduce~Eq.~(35) if $\omegabar\rightarrow$ 0.  For both this and the
ultrahigh $P_\phi$ cases, the rotational corrections begin at 
${\cal O}(\omegabar^2)$ because of the angular averaging.  However, 
corrections of ${\cal O}(\omegabar /\beta_0)$ may enter.

For moderate rotational velocities $\omegabar$ = $\Omega R/c\la$ 0.1, the 
azimuthally-averaged corrections are ${\cal O}(\omegabar^2).$  Corrections not
explicitly treated are: ${\cal O}(1/\beta_0),$ ${\cal O}(\omegabar /\beta_0),$
and ${\cal O}(\omegabar\sin\theta_0/\beta_0)$ (Fig.~\ref{fig10}).
The multiplicative rotational corrections to the magnetic field in the
energy arise as an effective electric field induced, as seen by an
inertial observer, by the rotation.  Purely rotational (centrifugal) 
corrections also enter.

\begin{figure}
\vspace{22pc}
\caption{
Phase space for  the ultrahigh $P_\phi$ case, with negligible $\eta .$
Effects of ${\cal O}(1/\beta_0)$ are greatly exaggerated.  (a) Phase space
for $\phi$ (degrees), showing small variation of $l_\phi$ for the case 
$\theta_0\neq$ 0.  The allowed region (phase space trajectory) itself has finite 
${\cal O}(1/\beta_0)$ thickness (not shown).  (b) Phase space for $\theta$
(degrees) showing small variation of $\theta$ and small values of $l_\theta 
.$}
\label{fig10}
\end{figure}

\section*{6. CONCLUSION}

We have examined only the simplified scenario of charged particles on a
rotating spherical surface with an intense dipole magnetic field.  The
resulting energy levels and magnetic flux quantization, for very high field 
strength, are roughly similar to the relativistic Landau and quantum Hall
systems.  The effect of realistic rotation velocities $(\Omega R/c\simeq$ 
0.1 or less) is small to moderate in the two-dimensional case.  For these
small values of $\omegabar$, the case of a tilted axis is not qualitatively 
different, although non-zero tilt creates a non-trivial geometry and 
complicates the dynamics.

There are a number of related issues in this idealized situation not treated
here.  A treatment of the three-dimensional orbitals requires inclusion of 
the radial motion and radial dependence of the vector potential ${\bf A}$
and has been considered for particles confined within
the crust in a separate work~\cite{gelb98}.
A fully quantum treatment of the orbitals necessitates
the Dirac equation and the inclusion of spin, although the basic features of
the Dirac orbitals are expected to be outlined by the W-S method.

Application of our results to a neutron star demands further realism.  In 
particular, the effect of the lattice of nuclei on non-localized electrons 
and nuclei must be included~\cite{shap83,rude75,helf87}.
The lattice structure of a neutron star crust 
must be disordered in its surface layers, although the fermion 
temperature is small compared to the Fermi energy~\cite{mich91}.  These lattice
effects modify the magnetosphere charge cloud,
the conduction bands, and the flux quantization.  The lattice must be
included to give a full picture of the crust's charge and current
distribution.  The magnetic field itself is also modified if the
currents contribute significantly to the field.

Finally, observational questions remain.  These include verification, if 
possible, of  flux quantization and determination of what roles it and the 
special conduction band structure play in the formation and evolution
of a neutron star's magnetic field and crust.\\

\section*{ACKNOWLEDGMENTS}

The authors are indebted to Sudheer Maremanda
(U.T. Arlington) for preparing the figures.  
We thank Pradeep Kumar (Univ. Florida)
for the original suggestion 
of this problem and Bernard Whiting (Univ. Florida) for helpful suggestions.  
This work was
supported at the Univ. of Florida
Institute for Fundamental Theory and U.S. 
Department of Energy Contract DE-FG05-86-ER40272;
and at the Univ. of Texas at Arlington by the 
Research Enhancement Program.


\begin{references}

\bibitem[*]{gelb}Email:~gelb@alum.mit.edu.

\bibitem[{\dagger}]{kennedy}Email:~kennedy@phys.ufl.edu.

\bibitem{long92} M. S. Longair, {\it High Energy Astrophysics}
(Cambridge University Press, Cambridge, 1992).
 
\bibitem{mich91} F. C. Michel, {\it Theory of Neutron Star Magnetospheres}
(University of Chicago Press, Chicago, 1991).

\bibitem{shap83} S. L. Shapiro and S. A. Teukolsky, {\it Black Holes, White 
Dwarfs, and Neutron Stars: The Physics of Compact Objects} (John Wiley 
\& Sons, New York, 1983).

\bibitem{frad91} E. Fradkin,
{\it Field Theories of Condensed Matter
Systems} (Addison-Wesley, Reading, 1991).

\bibitem{gelb98} J. M. Gelb, K. S. Gopinath, and D. C. Kennedy,
Univ. of Florida preprint UF-HEP-IFT-97-11 and Los Alamos
archive astro-ph/9707196, submitted to Phys. Rev. D.

\bibitem{lyne90} A. G. Lyne and F. Graham-Smith, {\it Pulsar Astronomy}
(Cambridge University Press, Cambridge, 1990).

\bibitem{rude75} M. A. Ruderman and P. G. Sutherland, Astrophys. J. 
{\bf 196}, 51 (1975).

\bibitem{land75} L. D. Landau and E. M. Lifshitz,
{\it The Classical Theory of
Fields} (Butterworth-Heinemann, Oxford, 1975).

\bibitem{land77} L. D. Landau and E. M. Lifshitz, {\it Quantum Mechanics:
Non-Relativistic Theory} 
(Pergamon Press, Oxford, 1977).

\bibitem{bere82} V. B. Berestetskii, 
E. M. Lifshitz, and L. P.
Pitaevskii,
{\it Quantum Electrodynamics} (Pergamon Press, Oxford, 1982)

\bibitem{land76} L. D. Landau and E. M. Lifshitz, {\it Mechanics}
(Pergamon Press, Oxford, 1976).

\bibitem{helf87} D. J. Helfand and J. H. Huang, in {\it Proc. IAU Symp. 125,
The Origin and 
Evolution of Neutron Stars} (Reidel, Dordrecht, 1987).

\end{references}
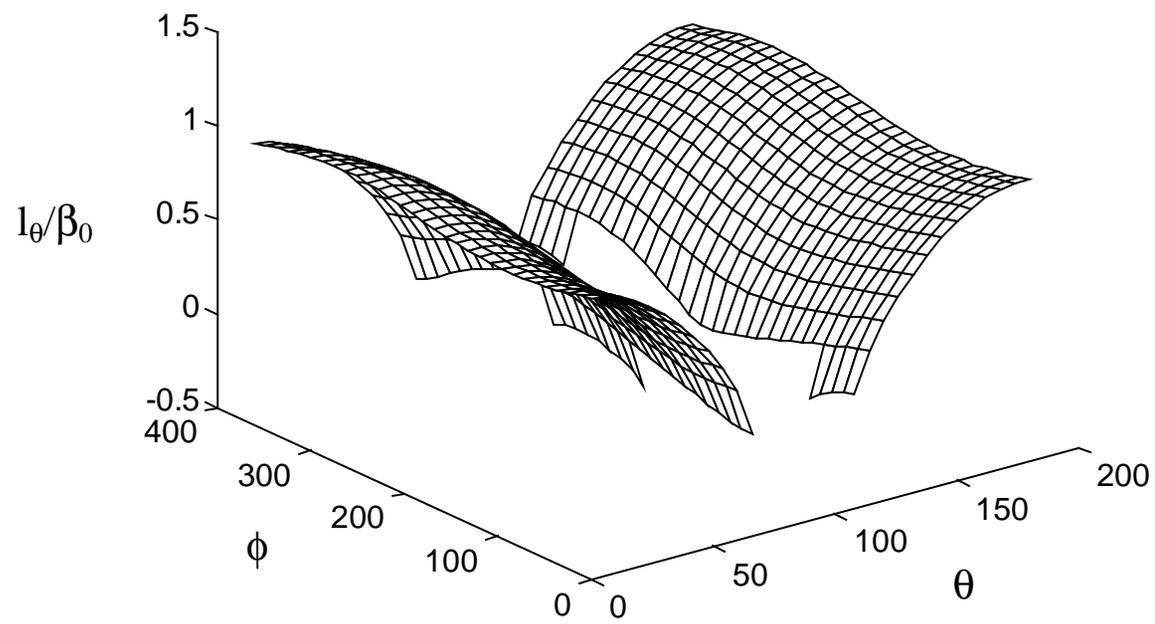
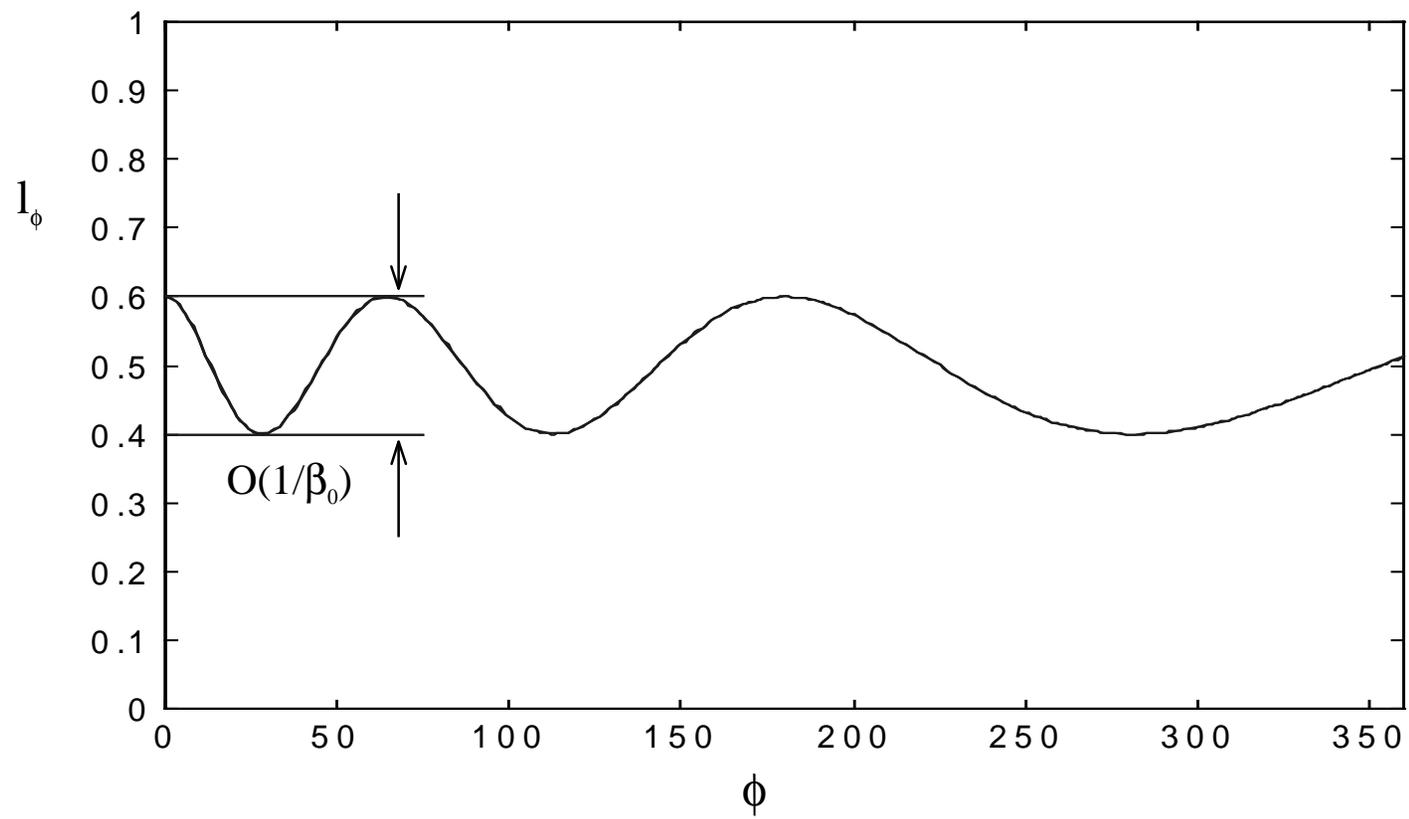
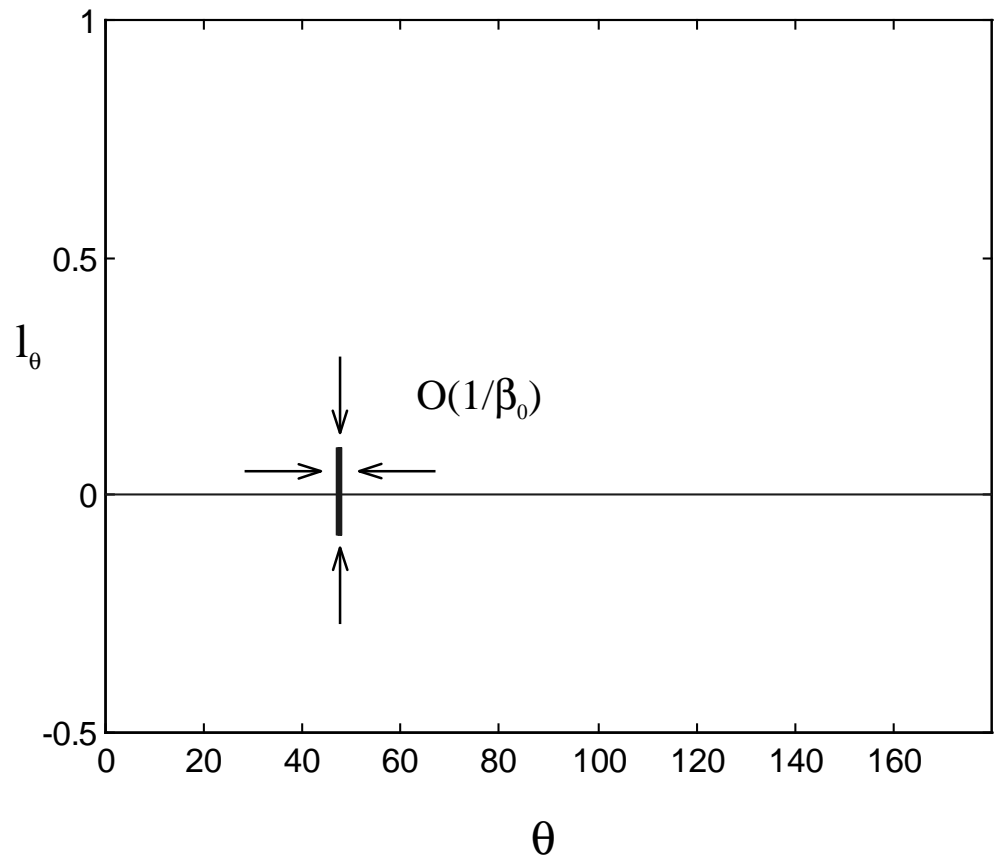
\end{document}